# Beyond Benchmarks: A Framework for Post-Deployment Validation of CT Lung Nodule Detection AI

Daniel Soliman

## Abstract

**Background:** Artificial intelligence (AI)-assisted lung nodule detection systems are increasingly deployed in clinical settings without site-specific validation. Performance reported under benchmark conditions may not reflect real-world behavior when acquisition parameters differ from training data.

**Purpose:** To propose and demonstrate a physics-guided framework for evaluating the sensitivity of a deployed lung nodule detection model to systematic variation in CT acquisition parameters.

**Methods:** Twenty-one cases from the publicly available LIDC-IDRI dataset were evaluated using a MONAI RetinaNet model pretrained on LUNA16 (fold 0, no fine-tuning). Five imaging conditions were tested: baseline, 25% dose reduction, 50% dose reduction, 3 mm slice thickness, and 5 mm slice thickness. Dose reduction was simulated via image-domain Gaussian noise; slice thickness via moving average along the z-axis. Detection sensitivity was computed at a confidence threshold of 0.5 with a 15 mm matching criterion.

**Results:** Baseline sensitivity was 45.2% (57/126 consensus nodules). Dose reduction produced slight degradation: 41.3% at 25% dose and 42.1% at 50% dose. The 5 mm slice thickness condition produced a marked drop to 26.2% — a 19 percentage point reduction representing a 42% relative decrease from baseline. This finding was consistent across confidence thresholds from 0.1 to 0.9. Per-case analysis revealed heterogeneous performance including two cases with complete detection failure at baseline.

**Conclusion:** Slice thickness represents a more fundamental constraint on AI detection performance than image noise under the conditions tested. The proposed framework is reproducible, requires no proprietary scanner data, and is designed to serve as the basis for ongoing post-deployment QA in resource-constrained environments.

## 1. Introduction

The rapid proliferation of Artificial Intelligence (AI)-assisted diagnostic software has given rise to a unique ecosystem where product development and validation often occur simultaneously. In 2023, the number of FDA-cleared AI-assisted solutions in radiology was roughly 650; today that number exceeds 1,000 [1]. Although modality and image data may differ, external validation studies of AI algorithms have demonstrated performance deficits between training datasets and external datasets, sometimes as high as 24%. This performance degradation does not appear to be specific to modality or population. In a 2022 systematic review, 70 of 86 (81%) deep learning algorithms were found to underperform on external datasets — 49% performed modestly worse and 24% showed substantial diminishment [2].

This performance gap is not incidental, nor is it easily characterized. Developing evaluation metrics that generalize across software types and local deployment environments has proven difficult precisely because medical imaging AI is deployed into conditions that differ substantially from training — varying scanner hardware, acquisition protocols, reconstruction algorithms, and heterogeneous patient populations. It is the structure of this underlying manifold — the acquisition parameters, scanner characteristics, and protocol variations that shape every image before a model ever sees it — that current benchmarks fail to adequately characterize. This paper proposes a physics-guided validation framework for characterizing the sensitivity of a deployed lung nodule detection model to systematic variation in CT acquisition parameters.

Regulatory pathways have enabled rapid market entry for many AI-assisted solutions, with the predominant 510(k) clearance route requiring demonstration of substantial equivalence to existing devices rather than prospective clinical validation — a dynamic likely to accelerate the pace at which new products reach market. Predictably, some organizations have stepped into the resulting vacuum. Stanford and the RSNA have each published guidance addressing aspects of this problem [3], and Suri et al. offer perhaps the most ambitious attempt at a structural solution: ROADMAP, a radiology ontology for AI datasets, models, and projects, designed to enable semantic interoperability and automated reasoning across heterogeneous imaging data [5]. It is a genuine opening salvo.

The limitations, however, are substantial. ROADMAP does not yet contend fully with the pace of regulatory change — the FDA and European bodies are actively revising expectations around model lifecycle management, performance drift, and real-world monitoring, and any static taxonomy risks obsolescence. More critically, the framework includes demographic descriptors relevant to fairness but does not enforce bias mitigation practices, a gap that emerging work on model cards has begun to address. Perhaps most fundamentally, without external regulatory pressure, adoption of any voluntary standard remains difficult to compel.

These considerations point to a gap that existing initiatives have not yet addressed: community hospitals, which lack the funding, physics infrastructure, and protocol standardization of academic centers, remain largely without practical tools for post-deployment model validation. It was this gap that motivated the present work. Using 21 cases from the publicly available LIDC-IDRI dataset [6] and a MONAI RetinaNet model pretrained on LUNA16 [7], detection sensitivity was evaluated across five imaging conditions: baseline, 25% dose reduction, 50% dose reduction, 3 mm slice thickness, and 5 mm slice thickness. Image degradation was applied using a physics-guided, image-domain approach designed to approximate realistic acquisition variability. Dose reduction produced modest sensitivity loss of approximately 4 percentage points, while 5 mm slice thickness resulted in a 19 percentage point decrease — a 42% relative reduction from baseline.

Although demonstrated here on a single model and dataset, this framework is intentionally generalizable; the physics-guided degradation process can in principle be applied to any pretrained CT detection model. Implementation requires only datasets with known acquisition parameters, without access to raw projection data or proprietary reconstruction pipelines. This makes deployment feasible even under constrained resources. More broadly, the methodology is designed not as a one-time experiment but as the basis for ongoing QA activities, running in parallel with routine physics quality control.

## 2. Methods

### 2.1 Dataset

This pilot utilized 21 anonymized scans from the Lung Image Database Consortium (LIDC) Image Database Resource Initiative (IDRI). These included cases 0001–0012, 0016, 0042, 0043, 0086, 0087, 0089, 0092, 0093, 0151. Case 0013 was discarded due to a 5-slice NIfTI volume yield unsuitable for inference. All cases were converted to NIfTI format (.nii.gz), which is the required file type for MONAI inference pipeline. Direction cosine matrices were confirmed as identity for all volumes used, validating the coordinate transformation used to match model detections to annotated nodule positions.

### 2.2 Nodule Consensus

Nodule consensus was defined using a minimum reader threshold of three, consistent with the LUNA16 benchmark convention [7]. This threshold was selected to reduce the influence of low-consensus annotations — those marked by fewer than three of four radiologists — on sensitivity estimates. Nodule centroid positions were derived from radiologist contour annotations available in the DICOM XML files. A total of 126 consensus nodules were identified across the 21 cases.

### 2.3 Model

The MONAI RetinaNet, pretrained on LUNA16, was used for this study. The model was not fine-tuned and fold 0 weights were used throughout[7]. This was done to simulate a model's performance outside of its initial training environment, as might be seen in a real post-deployment situation.

### 2.4 Imaging Conditions

Five imaging conditions were evaluated. These included: a baseline condition representing the original acquisition, two dose reduction conditions (25% and 50% of baseline dose), and two slice thickness conditions (3 mm and 5 mm). Dose reduction was simulated by adding Gaussian noise to the image volume, with standard deviation scaled as $\sigma \propto 1/\sqrt{\text{dose fraction}}$, consistent with established image-domain approximations of quantum noise in CT reconstruction [8, 9] — a viable alternative when raw projection data are unavailable.

Slice thickness increase was simulated via moving average along the z-axis. The 3 mm condition yielded an effective voxel spacing of 2.5 mm, which is noted here as a methodological limitation.

### 2.5 Detection and Matching

A confidence threshold of 0.5 was applied to model outputs. Detections were matched to consensus nodule positions using a 15 mm Euclidean distance criterion in world coordinates. Physical spacing in millimeters was derived from each volume's DICOM origin and voxel spacing. The 15 mm threshold follows the LUNA16 evaluation convention, chosen to accommodate inter-reader annotation variability [7]. Sensitivity was computed as the fraction of consensus nodules with at least one matched detection under each imaging condition.

## 3. Results

Sensitivity results across all five imaging conditions are presented in Table 1. The baseline sensitivity was 45.2% (57/126), reflecting performance of an unaltered, pretrained model applied to an external dataset — a level consistent with the deployment gap documented in the broader literature [2].

Dose reduction produced modest and roughly symmetric degradation. Sensitivity at 50% dose was 42.1% (53/126) and at 25% dose was 41.3% (52/126), representing a decline of approximately 3–4 percentage points from baseline. The near-equivalence of the two dose conditions is notable: halving the dose fraction produced little additional sensitivity loss, suggesting the model is relatively tolerant of image-domain noise at these levels.

Slice thickness had a substantially larger effect. The 3 mm condition (41.3%, 52/126) was comparable to the dose conditions, with minimal degradation over baseline. The 5 mm condition, however, produced a marked drop to 26.2% (33/126) — a 19 percentage point reduction representing a 42% relative decrease from baseline. This asymmetry between noise tolerance and resolution sensitivity is the primary finding of this study.

Two cases warrant individual note. LIDC-IDRI-0016 yielded 0/10 detections across all five conditions, including baseline. This was a complete model failure independent of acquisition parameter variation, illustrating that the deployment gap can manifest even before degradation is applied. LIDC-IDRI-0009 similarly produced 0/2 across all conditions; investigation confirmed that the model's high-confidence detections did not correspond spatially to annotated nodule positions, suggesting a localization rather than a detection failure.

*Table 1. Detection sensitivity across five CT acquisition conditions (21 cases, 126 consensus nodules, ≥3 readers, confidence threshold = 0.5, match distance = 15 mm).*

| Imaging Condition | Detected | Total Nodules | Sensitivity (%) |
|---|---|---|---|
| Baseline | 57 | 126 | 45.2% |
| 25% Dose | 52 | 126 | 41.3% |
| 50% Dose | 53 | 126 | 42.1% |
| 3mm Thick | 52 | 126 | 41.3% |
| **5mm Thick** | **33** | 126 | **26.2%** |

*pp = percentage points. The 5mm slice thickness condition (highlighted) produced the largest sensitivity reduction: −19 pp from baseline, representing a 42% relative decrease.*

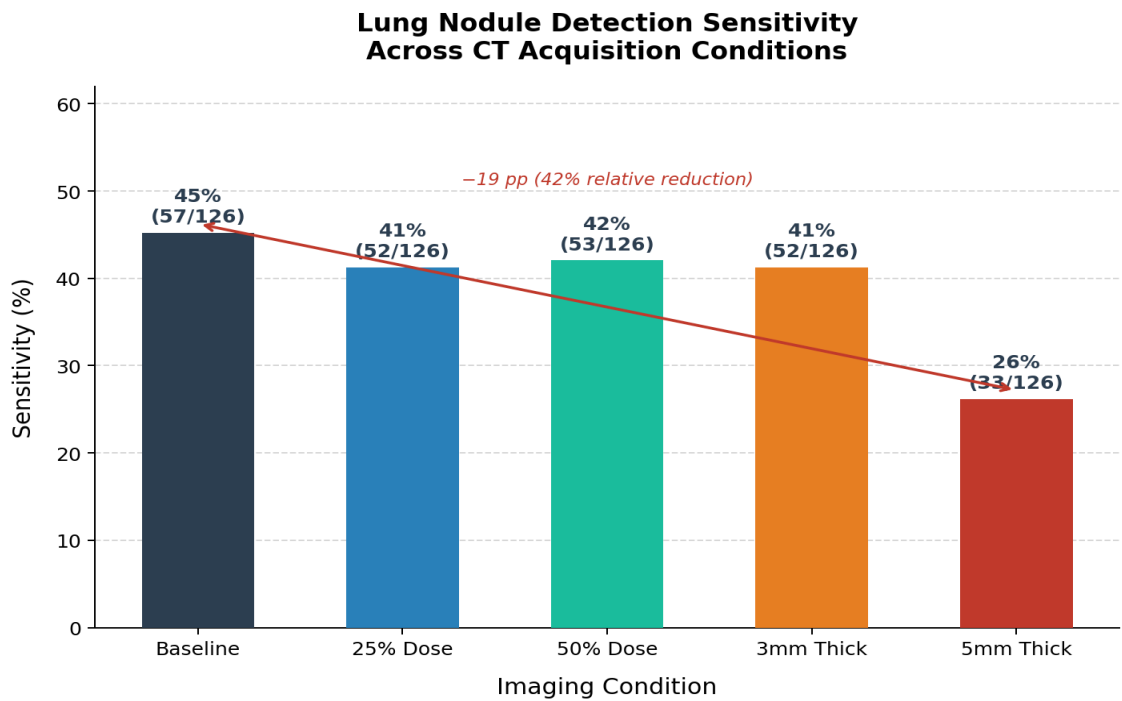

*Figure 1. Detection sensitivity by CT acquisition condition (21 cases, 126 consensus nodules).*

Figure 2 presents per-case detection results across all five conditions. Performance drift at the case level is notable; some cases, such as LIDC-IDRI-0005, maintain consistent sensitivity across all conditions, while others show marked variability. The factors underlying this heterogeneity may include nodule size, morphology, location, or acquisition characteristics — and warrant investigation in future work.

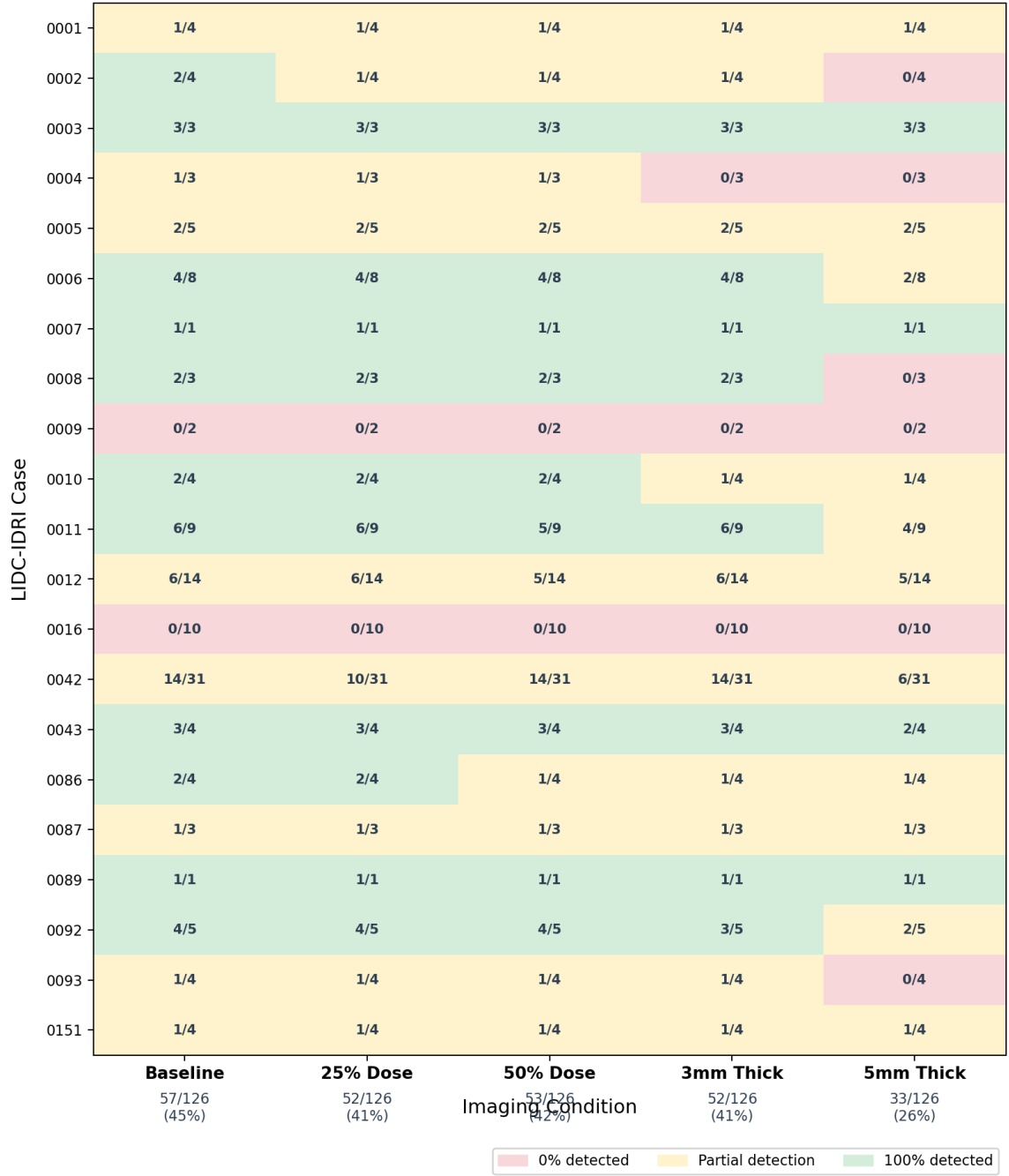

Figure 2: Per-case nodule detection fractions across five CT acquisition conditions. Color coding indicates detection rate per case: green (100%), yellow (partial), red (0%). Aggregate sensitivity is shown at column footers.

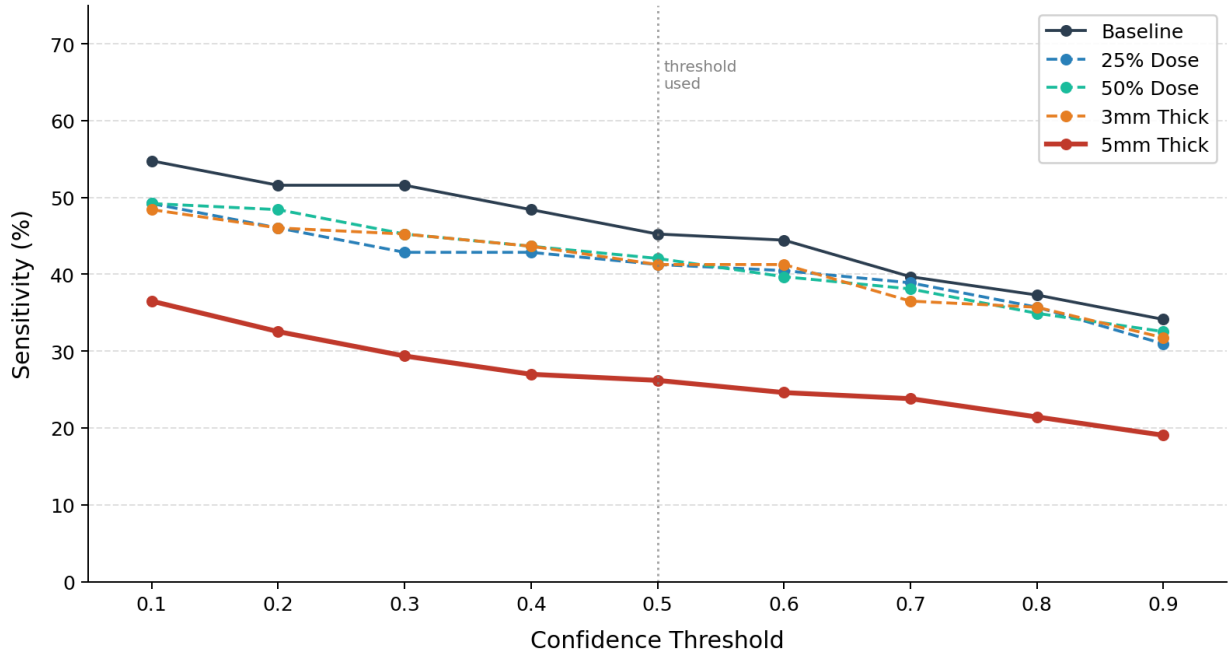

*Figure 3. Detection sensitivity as a function of confidence threshold across five CT acquisition conditions. The 5 mm slice thickness condition (red) remains consistently below all other conditions across the full threshold range (0.1–0.9), indicating that the observed sensitivity reduction is independent of threshold selection. Dose reduction conditions (blue tones) and 3 mm slice thickness (orange) track closely with baseline across all operating points. Vertical dotted line indicates the threshold used in the primary analysis (0.5)*

## 4. Discussion

The variable performance of the model provides an interesting starting point for expanded study using more cases. As this work represents a pilot to determine the feasibility of the method, there were some limitations. The case number (n = 21) was sufficient to analyze certain trends in the cases consistent with other work—namely that detection of models generally follows reader performance in that increased slice thickness reduces z-axis resolution.[10]

Perhaps more surprising is the heterogenous per-case performance. This is not easily explainable phenomenologically, as the cases showing absolute failure LIDC-IDRI-0016 and LIDC-IDRI-0009 — represent qualitatively distinct failure modes. In 0016, the model produced no detections whatsoever across all five conditions, including baseline, suggesting a case-level incompatibility between the model's learned feature representations and the imaging characteristics of this particular scan. In 0009, detections were produced with high confidence, but did not correspond spatially to annotated nodule positions. This would indicate a localization rather than a detection failure. Both cases illustrate that aggregate sensitivity metrics can obscure clinically meaningful variation in the nature of model failure — a finding that underscores the value of per-case audit frameworks.

*Robustness: Dose vs. Slice Thickness*

Model performance stayed relatively flat for dose reduction. This is very much in line with previous observation and is an illustration of robustness under noise. Because the added noise has both positive and negative perturbations (zero-mean), the expected value of each voxel intensity is preserved. This leaves spatial structure intact while degrading local signal-to-noise ratio: $E[I + \varepsilon] = E[I]$ when $\varepsilon \sim N(0, \sigma^2)$.[8]

Slice thickness increases, however, cause signal averaging across the z-axis, which reduces effective nodule contrast through the partial volume effect and distorts spatial resolution, as seen definitively in the 5 mm case [10].

*Baseline Sensitivity*

The publicly available MONAI RetinaNet implementation trained on LUNA16 was used as distributed, with fold 0 weights and without modification. It is hoped that this might function as a simulacrum of the practical reality that some if not most clinical deployments use off-the-shelf model configurations rather than optimized ensembles. [2]

*Limitations*

The cohort size of 21 cases represents the primary limitation of this work. A larger sample would be required to draw statistically robust conclusions about sensitivity estimates and their variance across cases. The image-domain noise approximation, while consistent with established methods [8, 9] does not fully capture the complexity of real dose reduction, which involves changes in projection-domain statistics, reconstruction kernel behavior, and scanner-specific noise texture. Modeling this characteristic completely would require access to specific scanner data. Finally, the use of a single model with a single fold does not represent the full spectrum of possible deployment environments — different architectures, training datasets, and reconstruction pipelines may respond differently to the acquisition parameter variations tested here.

## 5. Conclusion

This paper sought to provide a methodology for evaluating AI-assisted lung nodule detection sensitivity under systematic variation of CT acquisition parameters. Using the publicly available LIDC-IDRI dataset and MONAI RetinaNet model, it was demonstrated that dose reduction produced modest sensitivity loss of approximately 4 percentage points, while 5 mm slice thickness resulted in a 19 percentage point decrease — a finding that was consistent across confidence thresholds from 0.1 to 0.9. This asymmetry suggests that z-resolution represents a more fundamental constraint on detection performance than image noise, with implications for protocol design, patient dose reduction initiatives and model deployment at the facility level. The framework proposed here is reproducible, requires no proprietary data or scanner access, and is designed to serve as the basis for ongoing post-deployment QA. Future work should validate these findings across larger cohorts and multiple model architectures.